\pdfoutput=1

\documentclass[12pt,a4paper]{article}

\usepackage{ifthen} 
\newboolean{pdflatex}
\setboolean{pdflatex}{true} 

\newboolean{articletitles}
\setboolean{articletitles}{true} 

\newboolean{uprightparticles}
\setboolean{uprightparticles}{false} 

\newboolean{inbibliography}
\setboolean{inbibliography}{false} 

\newboolean{wordcount}
\setboolean{wordcount}{false} 

\usepackage{booktabs}

\usepackage[top=1in, bottom=1.25in, left=1in, right=1in]{geometry}

\usepackage{microtype}
\usepackage{lineno}  
\usepackage{xspace} 
\usepackage{caption} 

\usepackage{graphicx}  
\usepackage{color}
\usepackage{colortbl}
\graphicspath{{./figs/}} 

\usepackage{amsmath} 
\usepackage{amssymb}
\usepackage{amsfonts}
\usepackage{upgreek} 

\newcommand*\patchAmsMathEnvironmentForLineno[1]{%
\expandafter\let\csname old#1\expandafter\endcsname\csname #1\endcsname
\expandafter\let\csname oldend#1\expandafter\endcsname\csname
end#1\endcsname
 \renewenvironment{#1}%
   {\linenomath\csname old#1\endcsname}%
   {\csname oldend#1\endcsname\endlinenomath}%
}
\newcommand*\patchBothAmsMathEnvironmentsForLineno[1]{%
  \patchAmsMathEnvironmentForLineno{#1}%
  \patchAmsMathEnvironmentForLineno{#1*}%
}
\AtBeginDocument{%
\patchBothAmsMathEnvironmentsForLineno{equation}%
\patchBothAmsMathEnvironmentsForLineno{align}%
\patchBothAmsMathEnvironmentsForLineno{flalign}%
\patchBothAmsMathEnvironmentsForLineno{alignat}%
\patchBothAmsMathEnvironmentsForLineno{gather}%
\patchBothAmsMathEnvironmentsForLineno{multline}%
\patchBothAmsMathEnvironmentsForLineno{eqnarray}%
}

\usepackage{hyperref}    
\usepackage[all]{hypcap} 

\usepackage{xspace} 
\usepackage{upgreek}

\def\lhcb {\mbox{LHCb}\xspace}

\def\MagUp {\mbox{\em Mag\kern -0.05em Up}\xspace}

\ifthenelse{\boolean{uprightparticles}}%
{

 \def\Pmu         {\ensuremath{\upmu}\xspace}

 \def\Ppi         {\ensuremath{\uppi}\xspace}                 
                  
 \def\Prho        {\ensuremath{\uprho}\xspace}

 \def\Pomega      {\ensuremath{\upomega}\xspace}                 

 \def\PDelta      {\ensuremath{\Delta}\xspace}                 
 \def\PXi      {\ensuremath{\Xi}\xspace}                 
 \def\PLambda      {\ensuremath{\Lambda}\xspace}                 
 \def\PSigma      {\ensuremath{\Sigma}\xspace}                 
 \def\POmega      {\ensuremath{\Omega}\xspace}                 
 \def\PUpsilon      {\ensuremath{\Upsilon}\xspace}                 
 
 \def\PB      {\ensuremath{\mathrm{B}}\xspace}                 
                  
 \def\PD      {\ensuremath{\mathrm{D}}\xspace}

 \def\PK      {\ensuremath{\mathrm{K}}\xspace}

 \def\Pb      {\ensuremath{\mathrm{b}}\xspace}                 
 \def\Pc      {\ensuremath{\mathrm{c}}\xspace}

 \def\Pi      {\ensuremath{\mathrm{i}}\xspace}

 \def\Pp      {\ensuremath{\mathrm{p}}\xspace}

 \def\Pu      {\ensuremath{\mathrm{u}}\xspace}

}
{

 \def\Pmu         {\ensuremath{\mu}\xspace}

 \def\Ppi         {\ensuremath{\pi}\xspace}                 
                  
 \def\Prho        {\ensuremath{\rho}\xspace}

 \def\Pomega      {\ensuremath{\omega}\xspace}                 
 \mathchardef\PDelta="7101
 \mathchardef\PXi="7104
 \mathchardef\PLambda="7103
 \mathchardef\PSigma="7106
 \mathchardef\POmega="710A
 \mathchardef\PUpsilon="7107
                  
 \def\PB      {\ensuremath{B}\xspace}                 
                  
 \def\PD      {\ensuremath{D}\xspace}

 \def\PK      {\ensuremath{K}\xspace}

 \def\Pb      {\ensuremath{b}\xspace}                 
 \def\Pc      {\ensuremath{c}\xspace}

 \def\Pi      {\ensuremath{i}\xspace}

 \def\Pp      {\ensuremath{p}\xspace}

 \def\Pu      {\ensuremath{u}\xspace}

}

\makeatletter
\ifcase \@ptsize \relax
  \newcommand{\miniscule}{\@setfontsize\miniscule{4}{5}}
\or
  \newcommand{\miniscule}{\@setfontsize\miniscule{5}{6}}
\or
  \newcommand{\miniscule}{\@setfontsize\miniscule{5}{6}}
\fi
\makeatother

\DeclareRobustCommand{\optbar}[1]{\shortstack{{\miniscule (\rule[.5ex]{1.25em}{.18mm})}
  \\ [-.7ex] $#1$}}

\def\mup        {{\ensuremath{\Pmu^+}}\xspace}
\def\mun        {{\ensuremath{\Pmu^-}}\xspace} 

\def\uquark    {{\ensuremath{\Pu}}\xspace}

\def\cquark    {{\ensuremath{\Pc}}\xspace}

\def\bquark    {{\ensuremath{\Pb}}\xspace}

\def\pion   {{\ensuremath{\Ppi}}\xspace}

\def\pip    {{\ensuremath{\pion^+}}\xspace}
\def\pim    {{\ensuremath{\pion^-}}\xspace}

\def\rhomeson {{\ensuremath{\Prho}}\xspace}
\def\rhoz     {{\ensuremath{\rhomeson^0}}\xspace}

\def\kaon    {{\ensuremath{\PK}}\xspace}
  \def\Kbar    {{\kern 0.2em\overline{\kern -0.2em \PK}{}}\xspace}

\def\KorKbar    {\kern 0.18em\optbar{\kern -0.18em K}{}\xspace}

\def\Kp      {{\ensuremath{\kaon^+}}\xspace}
\def\Km      {{\ensuremath{\kaon^-}}\xspace}

\newcommand{\omegaz}{\ensuremath{\Pomega}\xspace}

  \def\Dbar    {{\kern 0.2em\overline{\kern -0.2em \PD}{}}\xspace}
\def\D       {{\ensuremath{\PD}}\xspace}

\def\DorDbar    {\kern 0.18em\optbar{\kern -0.18em D}{}\xspace}
\def\Dz      {{\ensuremath{\D^0}}\xspace}
\def\Dzb     {{\ensuremath{\Dbar{}^0}}\xspace}

\def\Dstar   {{\ensuremath{\D^*}}\xspace}

\def\Dstarp  {{\ensuremath{\D^{*+}}}\xspace}
\def\Dstarm  {{\ensuremath{\D^{*-}}}\xspace}

\def\Bbar    {{\ensuremath{\kern 0.18em\overline{\kern -0.18em \PB}{}}}\xspace}

\def\BorBbar    {\kern 0.18em\optbar{\kern -0.18em B}{}\xspace}

  \def\Y#1S{\ensuremath{\PUpsilon{(#1S)}}\xspace}

\def\proton      {{\ensuremath{\Pp}}\xspace}

\def\Lbar        {{\ensuremath{\kern 0.1em\overline{\kern -0.1em\PLambda}}}\xspace}
\def\LorLbar    {\kern 0.18em\optbar{\kern -0.18em \PLambda}{}\xspace}

\def\to                 {\ensuremath{\rightarrow}\xspace}

\def\CP                {{\ensuremath{C\!P}}\xspace}

\newcommand{\dm}{{\ensuremath{\Delta m}}\xspace}

\def\AT#1     {\ensuremath{A_{\mathrm{T}}^{#1}}\xspace}           

\def\C#1      {\ensuremath{\mathcal{C}_{#1}}\xspace}                       
\def\Cp#1     {\ensuremath{\mathcal{C}_{#1}^{'}}\xspace}                    
\def\Ceff#1   {\ensuremath{\mathcal{C}_{#1}^{\mathrm{(eff)}}}\xspace}        
\def\Cpeff#1  {\ensuremath{\mathcal{C}_{#1}^{'\mathrm{(eff)}}}\xspace}       
\def\Ope#1    {\ensuremath{\mathcal{O}_{#1}}\xspace}                       
\def\Opep#1   {\ensuremath{\mathcal{O}_{#1}^{'}}\xspace}                    

\newcommand{\tev}{\ifthenelse{\boolean{inbibliography}}{\ensuremath{~T\kern -0.05em eV}}{\ensuremath{\mathrm{\,Te\kern -0.1em V}}}\xspace}
\newcommand{\gev}{\ensuremath{\mathrm{\,Ge\kern -0.1em V}}\xspace}
\newcommand{\mev}{\ensuremath{\mathrm{\,Me\kern -0.1em V}}\xspace}
\newcommand{\kev}{\ensuremath{\mathrm{\,ke\kern -0.1em V}}\xspace}
\newcommand{\ev}{\ensuremath{\mathrm{\,e\kern -0.1em V}}\xspace}
\newcommand{\gevc}{\ensuremath{{\mathrm{\,Ge\kern -0.1em V\!/}c}}\xspace}
\newcommand{\mevc}{\ensuremath{{\mathrm{\,Me\kern -0.1em V\!/}c}}\xspace}
\newcommand{\gevcc}{\ensuremath{{\mathrm{\,Ge\kern -0.1em V\!/}c^2}}\xspace}
\newcommand{\gevgevcccc}{\ensuremath{{\mathrm{\,Ge\kern -0.1em V^2\!/}c^4}}\xspace}
\newcommand{\mevcc}{\ensuremath{{\mathrm{\,Me\kern -0.1em V\!/}c^2}}\xspace}

\def\invfb   {\ensuremath{\mbox{\,fb}^{-1}}\xspace}

\def\gsim{{~\raise.15em\hbox{$>$}\kern-.85em
          \lower.35em\hbox{$\sim$}~}\xspace}
\def\lsim{{~\raise.15em\hbox{$<$}\kern-.85em
          \lower.35em\hbox{$\sim$}~}\xspace}

\def\pt         {\mbox{$p_{\mathrm{ T}}$}\xspace}

\def\tell1  {TELL1\xspace}
\def\ukl1   {UKL1\xspace}

\usepackage{cite} 
\usepackage{mciteplus}

\usepackage{color}

\newcommand{\hhp}{\ensuremath{h^+h^{({\mkern-1mu\prime})-}}\xspace}
\newcommand{\hhmm}{\ensuremath{h^+h^-\mu^+\mu^-}\xspace}

\newcommand{\Dhhmm}{\mbox{\ensuremath{\Dz\to h^+h^-\mu^+\mu^-}}\xspace}
\newcommand{\Dkkmm}{\mbox{\ensuremath{\Dz\to\Kp\Km\mu^+\mu^-}}\xspace}
\newcommand{\Dppmm}{\mbox{\ensuremath{\Dz\to\pip\pim\mu^+\mu^-}}\xspace}

\newcommand{\mypaperversion}{}
\newcommand{\mydate}{June 28, 2018}
\newcommand{\mytitle}{Measurement of angular and \CP asymmetries in \Dppmm and \Dkkmm decays}
\newcommand{\mylhcbpapernumber}{LHCb-PAPER-2018-020}
\newcommand{\mycernpapernumber}{CERN-EP-2018-162}

\newcommand{\Acpppmm}{\ensuremath{4.9}\xspace}
\newcommand{\AcpppmmStat}{\ensuremath{3.8}\xspace}
\newcommand{\AcpppmmSyst}{\ensuremath{0.7}\xspace}
\newcommand{\Afbppmm}{\ensuremath{3.3}\xspace}
\newcommand{\AfbppmmStat}{\ensuremath{3.7}\xspace}
\newcommand{\AfbppmmSyst}{\ensuremath{0.6}\xspace}
\newcommand{\Aphippmm}{\ensuremath{-0.6}\xspace}
\newcommand{\AphippmmStat}{\ensuremath{3.7}\xspace}
\newcommand{\AphippmmSyst}{\ensuremath{0.6}\xspace}
\newcommand{\Acpkkmm}{\ensuremath{0}\xspace}
\newcommand{\AcpkkmmStat}{\ensuremath{11}\xspace}
\newcommand{\AcpkkmmSyst}{\ensuremath{2}\xspace}
\newcommand{\Afbkkmm}{\ensuremath{0}\xspace}
\newcommand{\AfbkkmmStat}{\ensuremath{11}\xspace}
\newcommand{\AfbkkmmSyst}{\ensuremath{2}\xspace}
\newcommand{\Aphikkmm}{\ensuremath{9}\xspace}
\newcommand{\AphikkmmStat}{\ensuremath{11}\xspace}
\newcommand{\AphikkmmSyst}{\ensuremath{1}\xspace}

\newcommand{\mmumu}{\ensuremath{m(\mup\mun)}\xspace}
\newcommand{\mhh}{\ensuremath{m(h^+h^-)}\xspace}
\newcommand{\mD}{\ensuremath{m(\hhmm)}\xspace}
\newcommand{\Acp}{\ensuremath{A_{\CP}}\xspace}

\newcommand{\Araw}{\ensuremath{A_{\CP}^{\rm raw}}\xspace}
\newcommand{\Afb}{\ensuremath{A_{\mathrm{FB}}}\xspace}
\newcommand{\Aphi}{\ensuremath{A_{2\phi}}\xspace}

\newcommand{\integratedresults}[1]{
\begin{align#1}
\Afb(\Dppmm) &= (\phantom{-}\Afbppmm\pm\AfbppmmStat\pm\AfbppmmSyst)\%,\\
\Aphi(\Dppmm)&= (\Aphippmm\pm\AphippmmStat\pm\AphippmmSyst)\%,\\
\Acp(\Dppmm) &= (\phantom{-}\Acpppmm\pm\AcpppmmStat\pm\AcpppmmSyst)\%,\\
\Afb(\Dkkmm) &= (\Afbkkmm\pm\AfbkkmmStat\pm\AfbkkmmSyst)\%,\\
\Aphi(\Dkkmm)&= (\Aphikkmm\pm\AphikkmmStat\pm\AphikkmmSyst)\%,\\
\Acp(\Dkkmm) &= (\Acpkkmm\pm\AcpkkmmStat\pm\AcpkkmmSyst)\%,
\end{align#1}
}

\def\paperauthors{LHCb collaboration} 
\def\papertitle{\mytitle} 
\def\papercopyright{\the\year\ CERN for the benefit of the LHCb collaboration} 
\def\paperlicence{CC-BY-4.0 licence}
\def\paperlicenceurl{https://creativecommons.org/licenses/by/4.0/}

\begin{document}
\renewcommand{\thefootnote}{\fnsymbol{footnote}}
\setcounter{footnote}{1}


\begin{titlepage}
\pagenumbering{roman}

\vspace*{-1.5cm}
\centerline{\large EUROPEAN ORGANIZATION FOR NUCLEAR RESEARCH (CERN)}
\vspace*{1.5cm}
\noindent
\begin{tabular*}{\linewidth}{lc@{\extracolsep{\fill}}r@{\extracolsep{0pt}}}
\ifthenelse{\boolean{pdflatex}}
{\vspace*{-1.5cm}\mbox{\!\!\!\includegraphics[width=.14\textwidth]{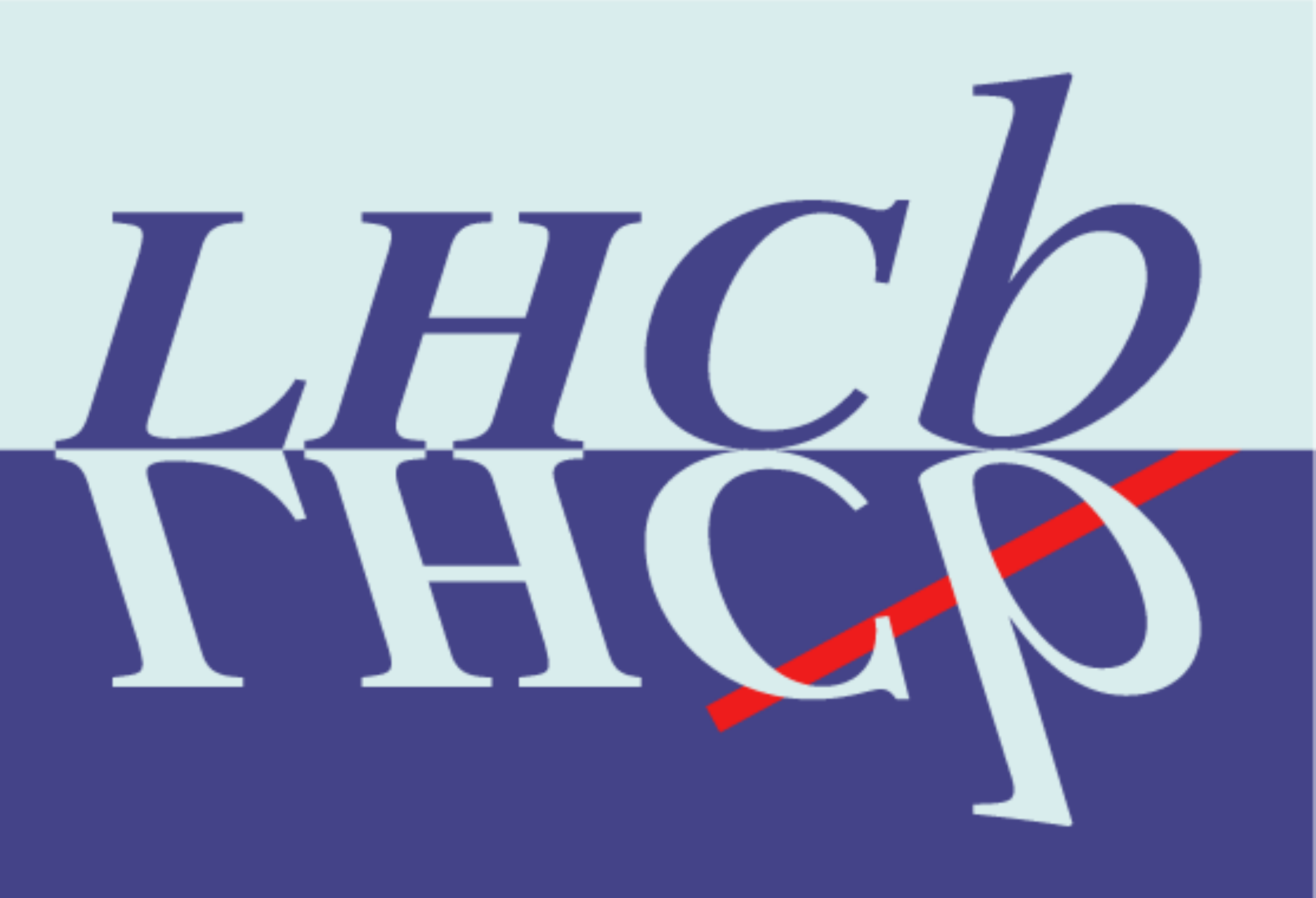}} & &}%
{\vspace*{-1.2cm}\mbox{\!\!\!\includegraphics[width=.12\textwidth]{lhcb-logo.eps}} & &}%
\\
 & & \mycernpapernumber \\  
 & & \mylhcbpapernumber \\  
 & & \mydate \\ 
 & & \mypaperversion \\
\end{tabular*}

\vspace*{3.0cm}

{\normalfont\bfseries\boldmath\huge
\begin{center}
  \papertitle 
\end{center}
}

\vspace*{1.0cm}

\begin{center}
\paperauthors\footnote{Authors are listed at the end of this paper.}
\end{center}

\vspace{\fill}

\begin{abstract}
  \noindent
  \noindent The first measurements of the forward-backward asymmetry of the dimuon pair (\Afb), the triple-product asymmetry (\Aphi), and the charge-parity-conjugation asymmetry (\Acp), in \Dppmm and \Dkkmm decays are reported. They are performed using data from proton-proton collisions collected with the LHCb experiment from 2011 to 2016, corresponding to a total integrated luminosity of 5\invfb. The asymmetries are measured to be
\integratedresults{*}
where the first uncertainty is statistical and the second systematic. The asymmetries are also measured as a function of the dimuon invariant mass. The results are consistent with the Standard Model predictions.

\end{abstract}

\vspace*{1.0cm}

\begin{center}
  Published in Phys.~Rev.~Lett.\ {\bf 121} (2018) 091801.
\end{center}

\vspace{\fill}

{\footnotesize 
\centerline{\copyright~\papercopyright. \href{\paperlicenceurl}{\paperlicence}.}}
\vspace*{2mm}

\end{titlepage}


\newpage
\setcounter{page}{2}
\mbox{~}
%
%
%
%

\cleardoublepage

\renewcommand{\thefootnote}{\arabic{footnote}}
\setcounter{footnote}{0}


\pagestyle{plain} 
\setcounter{page}{1}
\pagenumbering{arabic}


Decays of charm hadrons into final states containing muon pairs may proceed via the so-called {\em short-distance} $\cquark\to\uquark\mup\mun$ flavor-changing neutral-current process. In the Standard Model (SM) such process can only occur through electroweak loop transitions that are highly suppressed by the Glashow-Iliopoulos-Maiani mechanism~\cite{GIM}. The short-distance contribution to the inclusive $\D\to X\mup\mun$ branching fraction, where $X$ represents one or more hadrons, is predicted to be $\mathcal{O}(10^{-9})$~\cite{PaulBigi:2011}. The branching fraction can be greatly enhanced if new particles are exchanged in the loop, making these decays interesting for searches for physics beyond the SM. However, the SM branching fraction can increase up to $\mathcal{O}(10^{-6})$~\cite{Fajfer:2007,PaulBigi:2011,Cappiello,deBoer:2018} because of {\em long-distance} contributions occurring through tree-level amplitudes involving intermediate resonances that subsequently decay into $\mu^+\mu^-$. Hence, the sensitivity to the short-distance amplitudes is greatest for dimuon masses away from the peaks of the resonances, although resonances populate the entire dimuon-mass spectrum due to their long tails. Additional discrimination between short- and long-distance contributions can be obtained by studying kinematic correlations between final-state particles of multibody decays and charge-parity (\CP) conjugation asymmetries. These asymmetries are predicted to be negligibly small in the SM but could be as large as $\mathcal{O}(1\%)$ in scenarios of physics beyond the SM~\cite{Fajfer:2005ke,Bigi:2012,Paul:2012ab,Fajfer:2012nr,Cappiello,Fajfer:2015mia,deBoer:2015boa,deBoer:2018}.

\begin{figure}[b]
\centering
\includegraphics[width=.5\textwidth]{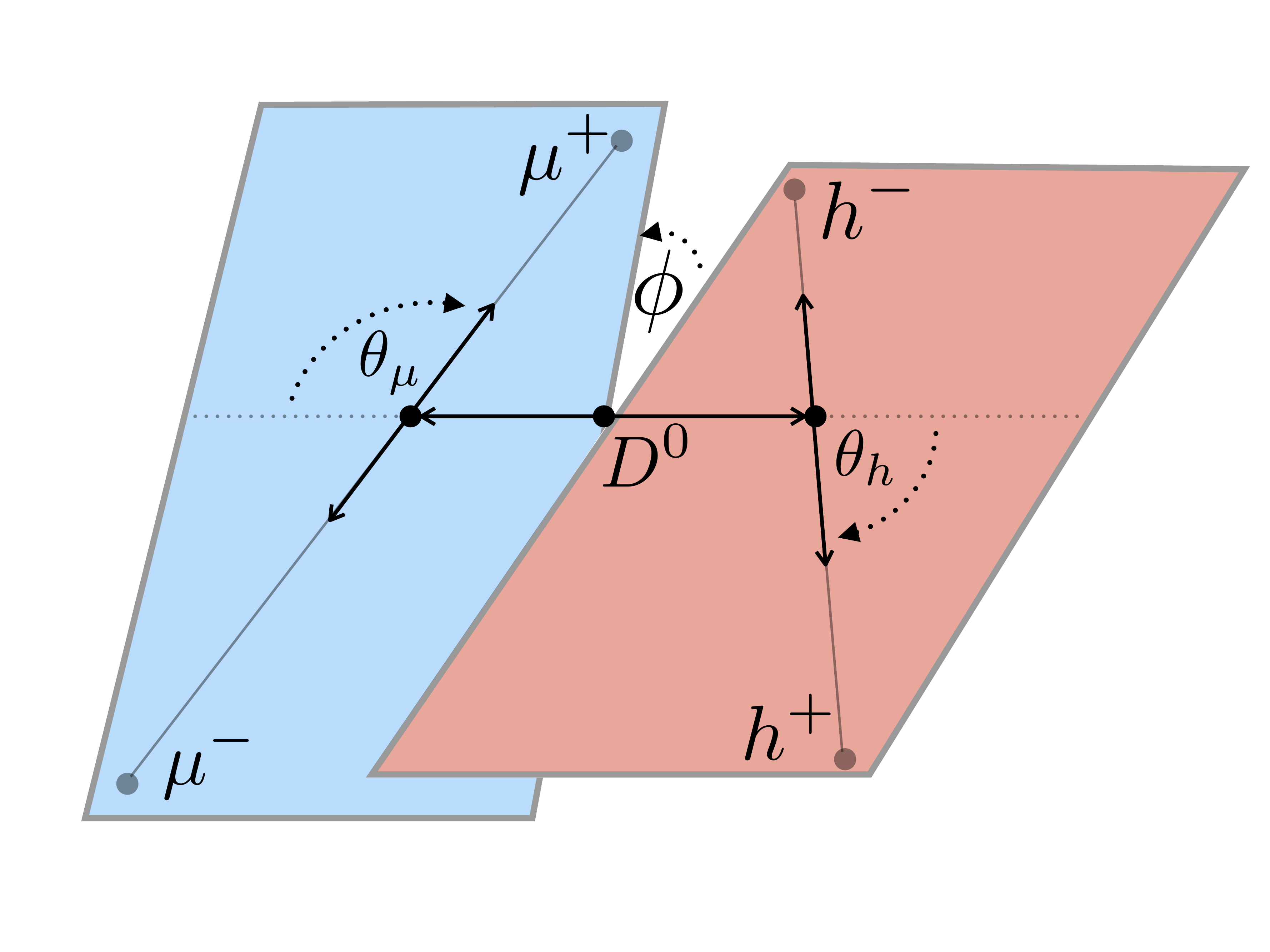}
\caption{Diagram showing the topology of a \Dhhmm decay, with the definition of the angles that are relevant for the measurement. 
}\label{fig:decayPlanes}
\end{figure}

The semileptonic four-body decays \Dhhmm (charge-conjugated decays are implied unless stated otherwise), where $h$ is either a pion or a kaon, are described by five independent kinematic variables (Fig.~\ref{fig:decayPlanes}): the dimuon invariant mass, \mmumu; the dihadron invariant mass, \mhh; the angle $\theta_\mu$ between the $\mu^+$ ($\mu^-$) direction and the direction opposite to the \Dz (\Dzb) meson in the dimuon rest frame; the angle $\theta_h$ between the $h^+$ ($h^-$) direction and the direction opposite to the \Dz (\Dzb) meson in the dihadron rest frame; and the angle $\phi$ between the two planes formed by the dimuon and the dihadron systems in the rest frame of the \Dz meson (the angle $\phi$ is zero if the two planes are parallel). Among all the possible angular observables that can be constructed, the forward-backward asymmetry of the dimuon system,
\ifthenelse{\boolean{wordcount}}{}{%
\begin{equation}
\Afb = \frac{\Gamma(\cos\theta_\mu>0)-\Gamma(\cos\theta_\mu<0)}{\Gamma(\cos\theta_\mu>0)+\Gamma(\cos\theta_\mu<0)},
\end{equation}}
and the triple-product asymmetry,
\ifthenelse{\boolean{wordcount}}{}{%
\begin{equation}
\Aphi = \frac{\Gamma(\sin2\phi>0)-\Gamma(\sin2\phi<0)}{\Gamma(\sin2\phi>0)+\Gamma(\sin2\phi<0)},
\end{equation}}
together with the \CP asymmetry,
\ifthenelse{\boolean{wordcount}}{}{%
\begin{equation}
\Acp = \frac{\Gamma(\Dz\to\hhmm)-\Gamma(\Dzb\to\hhmm)}{\Gamma(\Dz\to\hhmm)+\Gamma(\Dzb\to\hhmm)},
\end{equation}}
are considered to be promising probes for physics beyond the SM~\cite{Cappiello,deBoer:2018}.

The \Dppmm and \Dkkmm decays have been recently observed by the LHCb collaboration~\cite{LHCb-PAPER-2017-019} and their branching fractions have been measured to be $(9.6\pm1.2)\times 10^{-7}$ and $(1.54\pm0.33)\times 10^{-7}$, respectively, in agreement with SM predictions \cite{Cappiello,deBoer:2018}. However, angular and \CP asymmetries are yet to be measured.

This Letter reports the first measurement of \Afb, \Aphi and \Acp in \Dppmm and \Dkkmm decays using \proton\proton collision data collected with the \lhcb experiment at center-of-mass energies of 7, 8 and 13\tev between 2011 and 2016. The combined data set corresponds to a total integrated luminosity of 5\invfb. The analysis is performed using \Dz mesons originating from $\Dstarp\to\Dz\pip$ decays, with the \Dstarp meson produced at the primary \proton\proton collision vertex. The charge of the pion from the $\Dstar^\pm$ decay determines the flavor of the neutral \D meson at production. The signal is studied in regions of dimuon mass defined according to the known resonances as in Ref.~\cite{LHCb-PAPER-2017-019}. The regions dominated by the $\rhoz/\omega$ and the $\phi$ resonances are further split in two around the resonance mass to account for a possible variation of the asymmetries across the resonance~\cite{Fajfer:2012nr,deBoer:2015boa,deBoer:2018}. For \Dppmm decays the regions are: (low-mass) $<525\mevcc$, ($\eta$) $525$--$565\mevcc$, ($\rhoz/\omegaz$-low) $565$--$780\mevcc$, ($\rhoz/\omegaz$-high) $780$--$950\mevcc$, ($\phi$-low) $950$--$1020\mevcc$, ($\phi$-high) $1020$--$1100\mevcc$, and (high-mass) $>1100\mevcc$. The same regions are considered for \Dkkmm decays, except for the $\rhoz/\omega$ region, which is not split in two because of the reduced size of this sample, and the $\phi$ and high-mass regions, which are not kinematically accessible. The asymmetries are determined only in \mmumu regions where a significant signal yield was previously observed~\cite{LHCb-PAPER-2017-019}. No measurement is performed in the $\eta$ region of both channels and in the high-mass region of \Dppmm. Furthermore, to avoid potential experimenter's bias on the measured quantities, all asymmetries were shifted by an unknown value during the development of the analysis and examined only after the analysis procedure was finalized.

The \lhcb detector~\cite{Alves:2008zz,LHCb-DP-2014-002} is a single-arm forward spectrometer designed for the study of particles containing \bquark or \cquark quarks. It includes a high-precision tracking system consisting of a silicon-strip vertex detector surrounding the \proton\proton interaction region, a large-area silicon-strip detector located upstream of a dipole magnet with a bending power of about 4\,Tm, and three stations of silicon-strip detectors and straw drift tubes placed downstream of the magnet. The polarity of the magnetic field is reversed periodically throughout the data-taking. Particle identification is provided by two ring-imaging Cherenkov detectors, an electromagnetic and a hadronic calorimeter, and a muon system composed of alternating layers of iron and multiwire proportional chambers. Events are selected online by a trigger that consists of a hardware stage, which is based on information from the calorimeter and muon systems, followed by a software stage, based on information on charged tracks in the event that are displaced from any primary vertex. A subsequent software trigger exploits a full event reconstruction~\cite{LHCb-DP-2012-004} to exclusively select \Dhhmm decays. Candidate \Dz mesons are constructed by combining four charged tracks, each having momentum $p>3\gevc$ and transverse momentum $\pt>0.5\gevc$, that form a secondary vertex separated from any primary vertex in the event. Two oppositely charged particles are required to be identified as muons. 

The \Dz candidates satisfying the trigger requirements are further selected through particle-identification criteria placed on their decay products. Candidates with an invariant mass \mD in the range $1810$--$1940\mevcc$ are then combined with a charged particle originating from the same primary vertex and having $\pt>120\mevc$, hereafter referred to as soft pion, to form a $\Dstarp\to\Dz\pip$ decay candidate. When more than one primary vertex is reconstructed, the one with respect to which the \Dz candidate has the lowest impact-parameter significance is chosen. The vertex formed by the \Dz and \pip mesons is constrained to coincide with the primary vertex and the difference between the \Dstarp and \Dz masses, \dm, is required to be in the range $144.5$--$146.5$\mevcc (corresponding to approximately $\pm3\sigma$ in mass resolution around the signal peak). A fiducial-region requirement is implemented to reduce instrumental charge asymmetries that can bias the \Acp measurement. The fiducial region restricts the soft-pion trajectory to be within a fully instrumented region of the detector, because particles near the edge of the detector acceptance have different efficiencies to be reconstructed depending on their charge and the magnet polarity~\cite{LHCb-PAPER-2011-023}.

A multivariate selection, based on a boosted decision tree (BDT)~\cite{Breiman,Roe} with gradient boosting~\cite{TMVA}, is then used to suppress background from combinations of charged particles not originating from a \Dstarp decay. The features used in the BDT to discriminate signal from this {\em combinatorial} background are: the momentum and transverse momentum of the soft pion, the smallest impact parameter of the \Dz decay products with respect to the primary vertex, the angle between the \Dz momentum and the vector connecting the primary and secondary vertices, the quality of the secondary vertex, its separation from the primary vertex, and a measure of the isolation of the \Dstarp candidate from other tracks in the event. The BDT is trained separately for \Dppmm and \Dkkmm decays, because of their different kinematic properties, using simulated decays as signal and data candidates with \mD between 1890 and 1940\mevcc as background. A detailed description of the LHCb simulation can be found in Refs.~\cite{LHCb-PROC-2010-056,LHCb-PROC-2011-006}. To minimize biases on the background classification, the training samples are further randomly split in two disjoint subsets. The classifier trained on one subset is applied to the other, and vice versa.

The largest source of specific background is due to the hadronic four-body decays $\Dz\to\pip\pim\pip\pim$ and $\Dz\to\Kp\Km\pip\pim$, where two pions are misidentified as muons. The misidentification occurs mainly when the pions decay in flight into a muon and an undetected neutrino. This background is suppressed by a multivariate muon-identification discriminant that combines the information from the Cherenkov detectors, the calorimeters and the muon system~\cite{LHCb-DP-2013-001,LHCb-DP-2018-001}. Thresholds on the BDT response and on the muon-identification discriminant are optimized simultaneously by minimizing the statistical uncertainty on the measured asymmetry, as determined in data from randomly tagged \Dz and \Dzb candidates. After selection, less than 1\% of the events contain multiple candidates that share final-state particles. In these events one candidate is selected at random. The final samples comprise $1326\pm45$ \Dppmm and $137\pm14$ \Dkkmm signal decays, as determined from fits to the \mD distributions.

The selected candidates are corrected for any distortion of the phase space caused by the reconstruction and selection criteria. The efficiency for reconstruction and selection is modeled across the full five-dimensional phase space. This is achieved by using a BDT with gradient boosting~\cite{Breiman,Roe,TMVA} as a classification tool that learns about the different features of the generated and selected samples and combines them into a single variable~\cite{Viaud2016}. A weight corresponding to the inverse of the per-candidate efficiency is then computed as the ratio between the predicted probabilities for selected and generated candidates as a function of the BDT response. The training of this \textit{weighting} BDT is performed on simulated events before and after selection, using $|\cos\theta_\mu|$, $|\cos\theta_h|$, \mhh and \mmumu as input variables. This choice is justified by the observation that the efficiency is symmetric with respect to the angular variables, that it does not depend on $\sin 2\phi$, and that $\sin 2\phi$ is not correlated with any other variable. As a consequence of the efficiency weighting, the effective sample size of the \Dppmm (\Dkkmm) sample is reduced by about 13\% (14\%).

The \CP-asymmetry measurement is affected by $\mathcal{O}(1\%)$ nuisance charge asymmetries introduced by the different efficiency to reconstruct a positively or negatively charged soft pion, $A_D(\pip)$, and the different production cross-sections of \Dstarp and \Dstarm mesons, $A_P(\Dstarp)$. For small asymmetries, the {\em raw} asymmetry between observed yields of \mbox{$\Dstarp\to\Dz(\to f)\pip$} and \mbox{$\Dstarm\to\Dzb(\to f)\pim$} decays, where $f$ is a \CP-symmetric final state, can be approximated as \mbox{$\Araw(f)\approx\Acp(f)+A_P(\Dstarp)+A_D(\pip)$}. The nuisance charge asymmetries are subtracted from the raw asymmetry using high-yield samples of \mbox{$\Dstarp\to\Dz(\rightarrow K^+K^-)\pip$} decays. Therefore, \CP asymmetry is given by \mbox{$\Acp(\hhmm) = \Araw(\hhmm) - \Araw(\Kp\Km) + \Acp(\Kp\Km)$}, where $\Acp(\Kp\Km)=(-0.06\pm0.18)\%$ is taken from the independent measurement of Ref.~\cite{LHCb-PAPER-2014-013}. To account for different kinematic distributions in the signal and control modes, the procedure is performed in disjoint ranges of transverse and longitudinal momentum of the \Dstarp candidate.

\begin{table*}[t]
\centering
\caption{Efficiency-weighted yields and measured signal asymmetries for (top) \Dppmm and (bottom) \Dkkmm decays in the dimuon-mass regions. For the asymmetries the first uncertainty is statistical and the second systematic. Measurements are reported only in regions where a significant signal was previously observed~\cite{LHCb-PAPER-2017-019}. The sum of the yields in the dimuon-mass regions is not expected to match the yield of the full range, because the latter includes also the regions where no yields are reported.}\label{table:yieldsandasy}
\resizebox{
\ifdim\width>\textwidth
	\textwidth
\else
	\width
\fi}{!}{
\ifthenelse{\boolean{wordcount}}{}{%
\begin{tabular}{r@{--}l r@{\,$\pm$\,}l r@{\,$\pm$\,}l r@{\,$\pm$\,}l r@{\,$\pm$\,}c@{\,$\pm$\,}l r@{\,$\pm$\,}c@{\,$\pm$\,}l r@{\,$\pm$\,}c@{\,$\pm$\,}l}
\toprule
\multicolumn{2}{c}{$\mmumu$} & \multicolumn{6}{c}{Efficiency-weighted yields} & \multicolumn{9}{c}{Signal asymmetries}\\
\multicolumn{2}{c}{[$\mevcc$]} & \multicolumn{2}{c}{Signal} & \multicolumn{2}{c}{Misid.~back.} & \multicolumn{2}{c}{Comb.~back.} & \multicolumn{3}{c}{$\Afb$ [\%]} & \multicolumn{3}{c}{$\Aphi$ [\%]} & \multicolumn{3}{c}{$\Acp$ [\%]} \\
\midrule
\multicolumn{17}{c}{$\Dppmm$} \\
\multicolumn{2}{c}{$<525$} & 90 & 17\hspace*{11.55pt} & \hspace*{7.2pt}233 & 25 & \hspace*{8.25pt}108 & 22 & 2 & 20 & 2 & $-28$ & 20 & 2\hspace*{19pt} & 17 & 20 & 2\hspace*{25pt} \\
525 & 565 & \multicolumn{2}{c}{$-$} & \multicolumn{2}{c}{$-$} & \multicolumn{2}{c}{$-$} & \multicolumn{3}{c}{$-$} & \multicolumn{3}{c}{$-$} & \multicolumn{3}{c}{$-$}\\
565 & 780 & 326 & 23 & 253 & 24 & 145 & 21 & 8.1 & 7.1 & 0.7 & 7.4 & 7.1 & 0.7 & $-12.9$ & 7.1 & 0.7\\
780 & 950 & 141 & 14 & 159 & 15 &  89 & 14 & 7 & 10 & 1 & $-14$ & 10 & 1 & 17 & 10 & 1\\
950 & 1020 & 244 & 16 & 63 & 13 & 43 & 9 & 3.1 & 6.5 & 0.6 & 1.2 & 6.4 & 0.5 & 7.5 & 6.5 & 0.7 \\
1020 & 1100 & 258 & 14 & 33 & 9 & 44 & 9 & 0.9 & 5.6 & 0.7 & 1.4 & 5.5 & 0.6 & 9.9 & 5.5 & 0.7 \\
\multicolumn{2}{c}{$>1100$} & \multicolumn{2}{c}{$-$} & \multicolumn{2}{c}{$-$} & \multicolumn{2}{c}{$-$} & \multicolumn{3}{c}{$-$} & \multicolumn{3}{c}{$-$} & \multicolumn{3}{c}{$-$}\\
\multicolumn{2}{c}{Full range}  & 1083 & 41 & 827 & 42 & 579 & 39 & \Afbppmm & \AfbppmmStat &\AfbppmmSyst & \Aphippmm & \AphippmmStat & \AphippmmSyst & \Acpppmm & \AcpppmmStat & \AcpppmmSyst\\
\midrule
\multicolumn{17}{c}{$\Dkkmm$} \\
\multicolumn{2}{c}{$<525$} & 32 & 8 & 5 & 13 & 124 & 20 & 13 & 26 & 4 & 9 & 26 & 3 & $-33$ & 26 & 4\\
525 & 565 & \multicolumn{2}{c}{$-$} & \multicolumn{2}{c}{$-$} & \multicolumn{2}{c}{$-$} & \multicolumn{3}{c}{$-$} & \multicolumn{3}{c}{$-$} & \multicolumn{3}{c}{$-$}\\
\multicolumn{2}{c}{$>565$} & 74 & 9 & 39 & 7 & 48 & 8 & 1 & 12 & 1 & 22 & 12 & 1 & 13 & 12 & 1\\
\multicolumn{2}{c}{Full range} & 110 & 13 & 49 & 12 & 181 & 19 & \Afbkkmm & \AfbkkmmStat & \AfbkkmmSyst & \Aphikkmm & \AphikkmmStat & \AphikkmmSyst & \Acpkkmm & \AcpkkmmStat & \AcpkkmmSyst \\
\bottomrule
\end{tabular}
}}
\end{table*}

\begin{figure}[t]
\centering
\includegraphics[width=.5\textwidth]{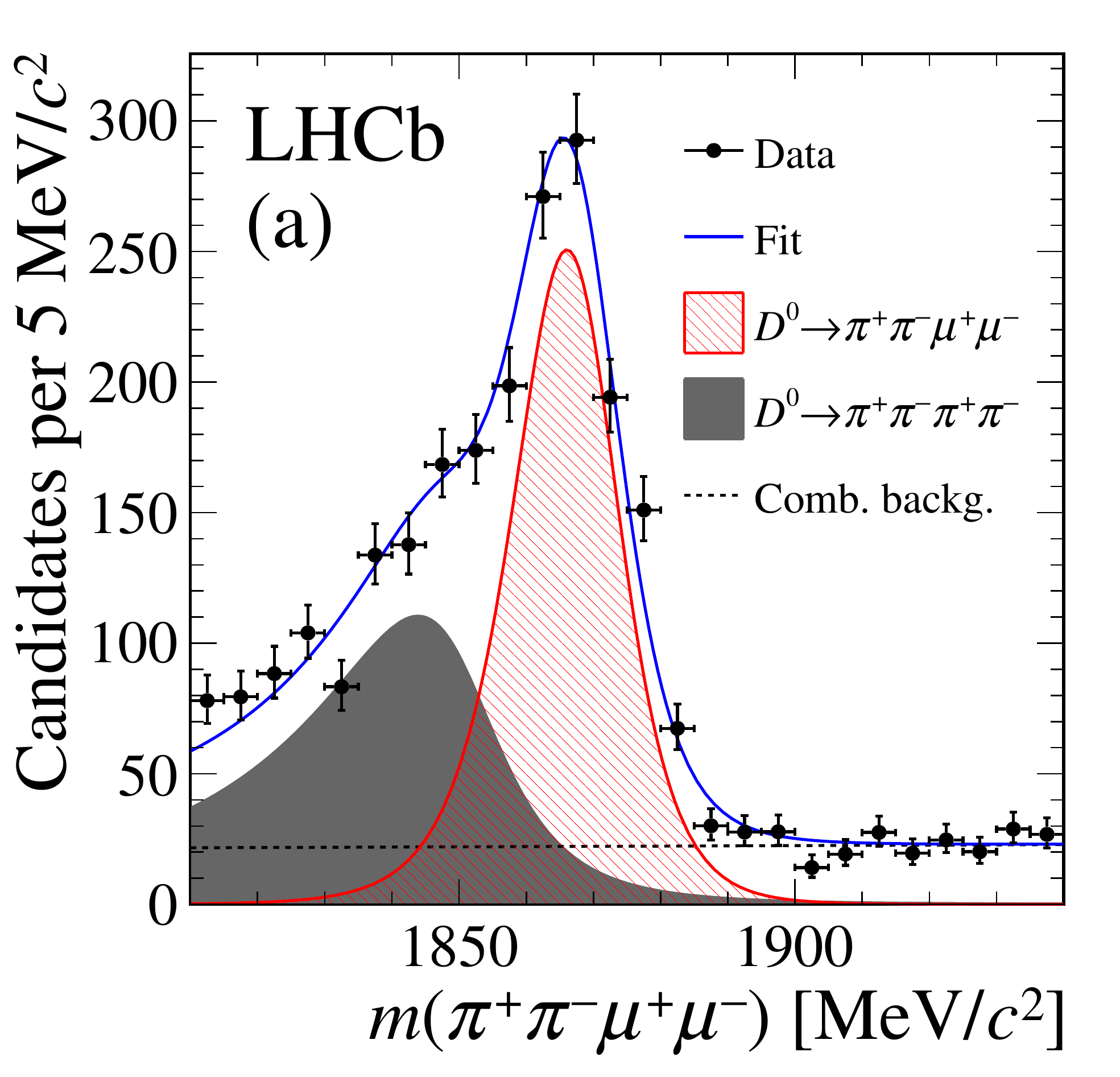}\hfil
\includegraphics[width=.5\textwidth]{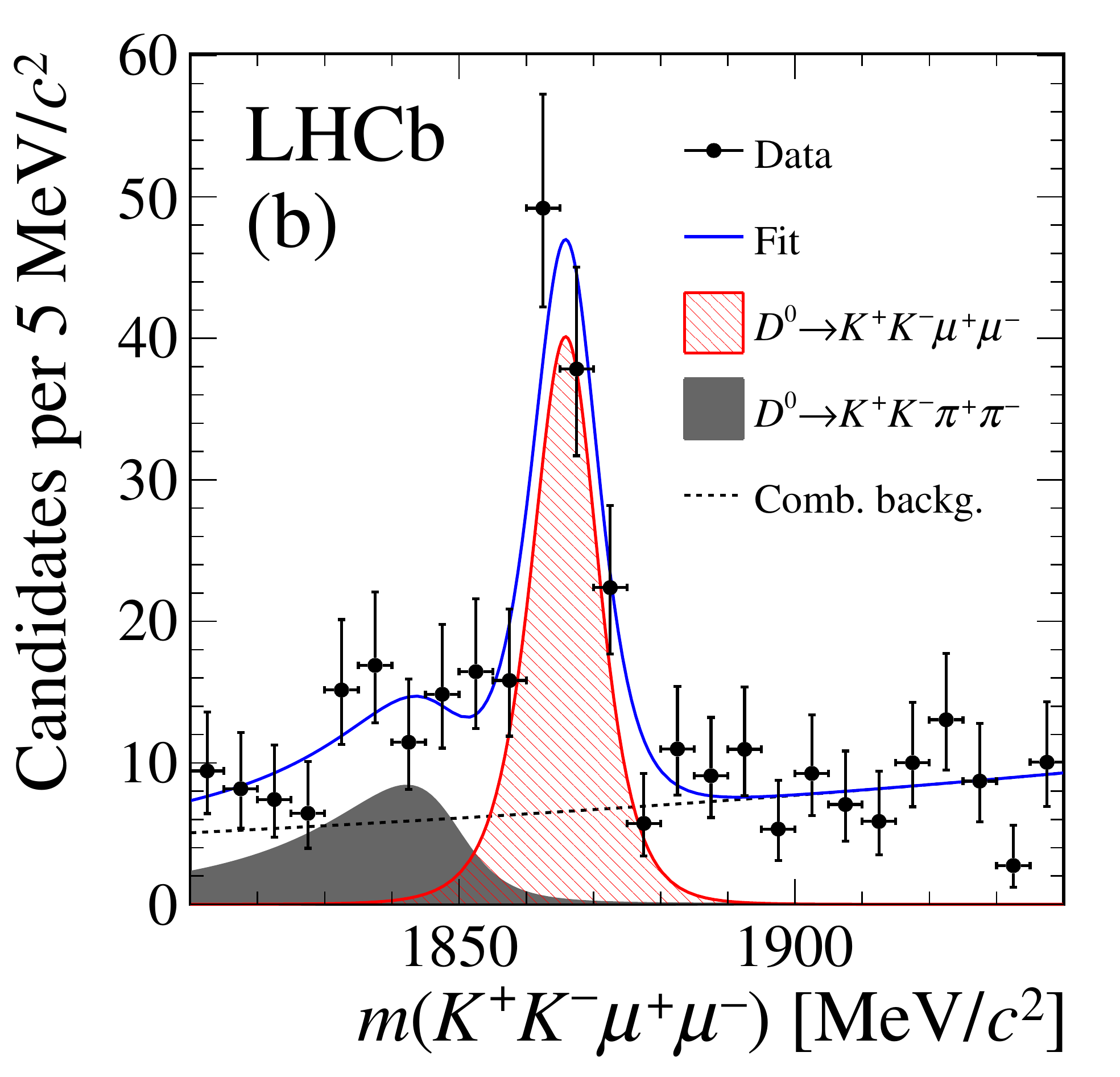}\\
\caption{Distribution of \mD for (a) \Dppmm and (b) \Dkkmm efficiency-weighted candidates, with fit projections overlaid.}\label{fig:FullDataFits}
\end{figure}

The asymmetries \Afb, \Aphi and \Acp of the signal decays are determined through unbinned maximum-likelihood fits to the $m(\hhmm)$ distributions of the selected candidates, weighted with the inverse of the per-candidate phase-space-dependent efficiency. The data are split into different tag categories (defined by the sign of $\cos{\theta_\mu}$, the sign of $\sin{2\phi}$ or the soft-pion charge) and a simultaneous fit is performed on the obtained data sets with the asymmetries as free parameters. The data are described by the sum of three components: the signal, the misidentified background and the combinatorial background. Analogously to Ref.{~\cite{LHCb-PAPER-2017-019}} the signal is described with a Johnson's $S_U$ distribution~\cite{johnson} with parameters determined from simulation. The mass shape of the misidentified background is determined using separate data samples of $\Dz\to\hhp\pip\pim$ decays where the \Dz mass is calculated assigning the muon-mass hypothesis to two oppositely charged pions. The combinatorial background is described by an exponential function. The shape of this background is fixed from data candidates with \dm above 150\mevcc that fail the BDT selection. Only the yields and the asymmetries of each component are allowed to vary in the fits, which are performed separately in each \mmumu region. The resulting efficiency-weighted yields are reported in Table~\ref{table:yieldsandasy}, together with the measured signal asymmetries. Figure~\ref{fig:FullDataFits} shows the \mD distribution of the efficiency-weighted candidates integrated in \mmumu, with the fit projection overlaid.

The following sources of systematic uncertainties affect the measured asymmetries: accuracy of the mass model used in the fit, accuracy of the phase-space-dependent efficiency, neglected asymmetric angular efficiencies and finite resolution on angular variables (affecting only \Afb and \Aphi); neglected background from \Dstarp candidates made up of correctly reconstructed \Dz candidates paired with unrelated soft pions (affecting only \Acp and \Afb); accuracy of the correction for the nuisance charge asymmetries and neglected backgrounds from \Dstarp candidates originating from $b$-hadron decays (affecting only \Acp). The leading systematic uncertainties are due to the accuracy of the efficiency correction, for all asymmetries; to possible asymmetric efficiencies as a function of $\cos\theta_\mu$ and $\sin2\phi$, for \Afb and \Aphi, respectively; and to the uncertainty on the nuisance charge asymmetry, for \Acp. The total systematic uncertainties amount to less than 20\% of the statistical uncertainties for both decay modes and all dimuon-mass regions (Table~\ref{table:yieldsandasy}).

The analysis is repeated on statistically independent data subsets chosen according to criteria likely to reveal biases from specific instrumental effects. These criteria include the data-taking year, the magnetic-field orientation, the number of primary vertices in the event, the per-event track multiplicity, the trigger classification, the \Dstarp transverse momentum and the impact parameter of the \Dz candidate with respect to the primary vertex. The resulting variations of the measured asymmetries are consistent with statistical fluctuations, with $p$-values between 3\% and 95\% and without deviations from a flat distribution.

In summary, measurements of angular and \CP asymmetries in \Dppmm and \Dkkmm decays are performed using the proton-proton collision data collected with the LHCb experiment between 2011 and 2016. This is the first time such measurements are performed with rare decays of charm hadrons. The asymmetries are measured both integrated and as a function of dimuon mass. The integrated asymmetries are
\ifthenelse{\boolean{wordcount}}{}{\integratedresults{*}}
where the first uncertainty is statistical and the second systematic. These measurements, as well as the asymmetries in each dimuon-mass region, are consistent with zero and will help constrain scenarios of physics beyond the SM~\cite{Fajfer:2005ke,Bigi:2012,Paul:2012ab,Cappiello,Fajfer:2015mia,deBoer:2015boa,deBoer:2018}.

\section*{Acknowledgements}
%
%
\noindent We express our gratitude to our colleagues in the CERN
accelerator departments for the excellent performance of the LHC. We
thank the technical and administrative staff at the LHCb
institutes. We acknowledge support from CERN and from the national
agencies: CAPES, CNPq, FAPERJ and FINEP (Brazil); MOST and NSFC
(China); CNRS/IN2P3 (France); BMBF, DFG and MPG (Germany); INFN
(Italy); NWO (Netherlands); MNiSW and NCN (Poland); MEN/IFA
(Romania); MinES and FASO (Russia); MinECo (Spain); SNSF and SER
(Switzerland); NASU (Ukraine); STFC (United Kingdom); NSF (USA).  We
acknowledge the computing resources that are provided by CERN, IN2P3
(France), KIT and DESY (Germany), INFN (Italy), SURF (Netherlands),
PIC (Spain), GridPP (United Kingdom), RRCKI and Yandex
LLC (Russia), CSCS (Switzerland), IFIN-HH (Romania), CBPF (Brazil),
PL-GRID (Poland) and OSC (USA). We are indebted to the communities
behind the multiple open-source software packages on which we depend.
Individual groups or members have received support from AvH Foundation
(Germany), EPLANET, Marie Sk\l{}odowska-Curie Actions and ERC
(European Union), ANR, Labex P2IO and OCEVU, and R\'{e}gion
Auvergne-Rh\^{o}ne-Alpes (France), Key Research Program of Frontier
Sciences of CAS, CAS PIFI, and the Thousand Talents Program (China),
RFBR, RSF and Yandex LLC (Russia), GVA, XuntaGal and GENCAT (Spain),
Herchel Smith Fund, the Royal Society, the English-Speaking Union and
the Leverhulme Trust (United Kingdom).

\addcontentsline{toc}{section}{References}
\setboolean{inbibliography}{true}
\bibliographystyle{LHCb}
\bibliography{main,LHCb-PAPER,LHCb-CONF,LHCb-DP,LHCb-TDR,mybib}

\ifthenelse{\boolean{wordcount}}{}{%
\newpage

\newpage
\centerline{\large\bf LHCb collaboration}
\begin{flushleft}
\small
R.~Aaij$^{27}$,
B.~Adeva$^{41}$,
M.~Adinolfi$^{48}$,
C.A.~Aidala$^{73}$,
Z.~Ajaltouni$^{5}$,
S.~Akar$^{59}$,
P.~Albicocco$^{18}$,
J.~Albrecht$^{10}$,
F.~Alessio$^{42}$,
M.~Alexander$^{53}$
% ...

{\footnotesize \it
% ...
$ ^{5}$Clermont Universit{\'e}, Universit{\'e} Blaise Pascal, CNRS/IN2P3, LPC, Clermont-Ferrand, France\\
% ...
$ ^{10}$Fakult{\"a}t Physik, Technische Universit{\"a}t Dortmund, Dortmund, Germany\\
% ...
$ ^{18}$INFN Laboratori Nazionali di Frascati, Frascati, Italy\\
% ...
$ ^{27}$Nikhef National Institute for Subatomic Physics, Amsterdam, Netherlands\\
% ...
$ ^{41}$Instituto Galego de F{\'\i}sica de Altas Enerx{\'\i}as (IGFAE), Universidade de Santiago de Compostela, Santiago de Compostela, Spain\\
$ ^{42}$European Organization for Nuclear Research (CERN), Geneva, Switzerland\\
% ...
$ ^{48}$H.H. Wills Physics Laboratory, University of Bristol, Bristol, United Kingdom\\
% ...
$ ^{53}$School of Physics and Astronomy, University of Glasgow, Glasgow, United Kingdom\\
% ...
$ ^{59}$University of Cincinnati, Cincinnati, OH, United States\\
% ...
$ ^{61}$Syracuse University, Syracuse, NY, United States\\
% ...
$ ^{73}$University of Michigan, Ann Arbor, United States, associated to $^{61}$\\
% ...
}
\end{flushleft}
}

\end{document}